\documentclass[12pt]{article}

\usepackage{scicite}

\usepackage{times}

\newenvironment{sciabstract}{%
\begin{quote} \bf}
{\end{quote}}

\newcounter{lastnote}
\newenvironment{scilastnote}{%
\setcounter{lastnote}{\value{enumiv}}%
\addtocounter{lastnote}{+1}%
\begin{list}%
{\arabic{lastnote}.} {\setlength{\leftmargin}{.22in}}
{\setlength{\labelsep}{.5em}}} {\end{list}}

\title{Universal Theory of Nonlinear Luttinger Liquids}

\author
{Adilet Imambekov,$^{1}$ Leonid I. Glazman,$^{1\ast}$ \\
\\
\normalsize{$^{1}$Department of Physics, Yale University,}\\
\normalsize{New Haven,
CT 06520, USA}\\
\\
\normalsize{$^\ast$To whom correspondence should be addressed;
E-mail:  Leonid.Glazman@yale.edu.} }

\date{}

\usepackage{graphicx}

\begin{document}




\maketitle


\begin{sciabstract}
One-dimensional quantum fluids are conventionally described by
using an effective hydrodynamic approach known as Luttinger liquid
theory.  As the principal  simplification, a generic spectrum of
the constituent particles is  replaced by a linear one, which
leads to a linear hydrodynamic theory. We show that to describe
the measurable  dynamic response functions one needs to take into
account the nonlinearity of the generic  spectrum  and thus of
the resulting quantum hydrodynamic theory. This nonlinearity
leads, for example, to a qualitative change in the behavior of
the spectral function.    The universal theory developed in this
article is applicable to a wide class of one-dimensional
fermionic, bosonic, and spin
  systems.
\end{sciabstract}

\maketitle

\def\be{\begin{equation}}
\def\ee{\end{equation}}
\def\bea{\begin{eqnarray}}
\def\eea{\end{eqnarray}}
\def\bean{\begin{mathletters}\begin{eqnarray}}
\def\eean{\end{eqnarray}\end{mathletters}}


The development of the universal effective description of
many-body phenomena is a central problem of the condensed matter
theory.
 The hydrodynamic approach known as Luttinger
liquid (LL) theory~\cite{Haldane_PRL,Haldane_long,Giamarchibook}
is routinely applied to one-dimensional (1D) interacting systems.
As a crucial simplification, a generic spectrum of the constituent
particles is replaced by a linear one, leading to a linear
hydrodynamic theory, which is nothing but a collection of
noninteracting oscillators. However, to understand a variety of
phenomena, such as Coulomb drag between quantum wires
\cite{drag}, momentum-resolved tunneling of electrons in
nanowires~\cite{Yacoby_science}, and neutron scattering off spin
chains~\cite{Giamarchibook}, one needs to take into account the
nonlinearity of the spectrum. From classical physics, it is known
that the existence of nonlinearities may result in qualitatively
new phenomena, such as propagation of solitons and appearance of
shock waves. These phenomena take place in a variety of
experimental situations because classical nonlinear hydrodynamics
is universal:  It is phenomenologically derived from simple
assumptions, which do not rely on microscopic details. Although
description of linear quantum hydrodynamic theory requires only
quantum mechanical treatment of noninteracting oscillators,
formulation of nonlinear quantum hydrodynamics remains a
challenging task because of divergences typical of nonlinear
quantum field theories. In this article, we develop a universal
theory of 1D quantum liquids that includes nonlinear hydrodynamic
effects, leading to qualitative changes in predictions for
dynamic response functions (e.g. spectral function).

If  the 1D quantum many-body problem  for fermions is simplified
by replacing a generic spectrum of particles by a linear one [the
Tomonaga-Luttinger (TL) model
\cite{Tomonaga,Luttinger,MattisLieb}], it becomes solvable at any
interaction strength. The Lorentz invariance introduced by this
simplification protects the existence of well-defined elementary
excitations with linear dispersion relation. These excitations
are  quantized waves of density propagating with a velocity  $v.$
 Adding a fermion to a 1D system described by the
  TL model requires creation of multiple elementary
  excitations \cite{DL,LP}. This can be seen from the form of the fermionic
  single-particle spectral function $A(p,\omega),$ which describes
  the probability of tunneling a fermion with given momentum $p$ and energy $\omega$ into the system [see
  Supporting Online Material (SOM) for the precise definition \cite{SOM}].
 The spectral function  has a power-law singularity at the energy of collective
excitation $\omega=vp$ (see Eq.~S4). The corresponding exponent is
determined only by the universal LL
  parameter $K$ [the latter is expressed in terms of the density, compressibility, and sound velocity $v$, the three low-energy properties of 1D liquid
 \cite{Giamarchibook}].

In the phenomenological LL approach, energy scale $p^2/(2m_*)$ is
fully dispensed with (the effective mass $m_*$ characterizes
spectrum nonlinearity at $p=0$ and is defined below).
 The conventional justification for such
simplification is irrelevance, in the renormalization group sense,
of the nonlinearity \cite{Haldane_long}. Indeed, the
  irrelevant terms hardly affect the fermion propagator
   away from the singular lines in space-time, $x \pm vt\gg\sqrt{t/m_*}$. However, it
  is the vicinity of these lines that defines the nature of singular
  behavior of the spectral function.
We show here that  for all  spinless 1D fermionic models with
short range interactions the single-particle spectral function at
$p\ll k_f$  is universal. In the vicinity of Fermi wave vector
$+k_f$ and $p,\omega>0,$ for example, $A(p,\omega)$ is  a
universal function of a single argument
 \bea A(p,\omega)\propto
 A(\varepsilon), \; \varepsilon=\frac{\omega- v
p}{p^2/2m_*} \label{edef} \eea (hereinafter $p$ is measured from
the closest Fermi point, and we use units with $\hbar=1$). The
new nontrivial function $A(\varepsilon)$ is very different from
the LL theory predictions, yet it depends only on the LL parameter
$K$. The asymptote of $A(\varepsilon)$ at $\varepsilon\gg 1$ does
reproduce the LL theory predictions, Eq.~S4, but at
$|\varepsilon\pm 1|\ll 1$ the spectral function is described by
power-law asymptotes with new exponents. The exponents are
different from the predictions of the LL theory but can still be
analytically expressed in terms of $K.$ We find numerically the
universal single-variable crossover function $A(\varepsilon),$ by
relating it to the nonlinear dynamics in nonequilibrium Fermi
gases \cite{AL1,AL2,AbanovWiegmannPRL2,AbanovWiegmannrecent}. We
also briefly discuss applications of our results for bosonic and
spin systems.

 The spectral function could be measured in tunneling
  experiments  with electrons in nanowires \cite{Yacoby_science}  and cold atoms in elongated
  traps~\cite{Duan, Carusotto}. A closely related object, transverse dynamic spin structure factor, is measurable by neutron scattering off
  1D  spin liquids placed in a magnetic field
~\cite{Giamarchibook}.
 The universal  crossover function and its analytically obtained asymptotes also
  provide one with a test for numerical methods  to evaluate many-body dynamics  of 1D models, for example,
using density-matrix renormalization group algorithms
~\cite{DMRG0,DMRG3}.

Within a LL approach, fermionic field $\Psi$ is expanded by using
its components near Fermi points as $ \Psi(x,t) \approx
\Psi^{\vphantom{\dagger}}_{\rm R}(x,t) e^{i k_f x} +
\Psi^{\vphantom{\dagger}}_{\rm L}(x,t) e^{-i k_f x},$ and the
kinetic energy term in the Hamiltonian is linearized. Solution of
the linearized model can be described by using free bosonic fields
with linear dispersion. Fermionic operators  are expressed as
exponentials of free bosonic fields, and their correlations are
easily evaluated. Including the nonlinearity of the spectrum of
constituent fermions leads to interactions between bosonic fields
\cite{Haldane_long,Pereira_short}. One cannot treat such
interactions perturbatively in bosonic language in the vicinity
of the line $\omega = v p$ because even in the second order of
perturbation theory corrections diverge  there \cite{Samokhin}.
Physically this happens because conservation laws of energy and
momentum are satisfied simultaneously for waves with linear
dispersion. Thus, two wave packets spend an infinite amount of
time near each other, leading to an ill-defined perturbation
theory. To understand the effects of nonlinear spectrum, it is
more convenient to work in the fermionic representation. Recently
a connection between dynamic response functions of 1D quantum
liquids and well-known Fermi edge singularity was
elucidated~\cite{Pustilnik2006Fermions,Khodas2006Fermions}. It
allowed one to evaluate dynamic structure factor  $S(p,\omega)$
and  spectral function perturbatively in the interaction between
fermions.  Moreover, it established the form of the effective
Hamiltonian defining the true  low-energy behavior of a liquid
composed of generic particles with nonlinear dispersion relation.
For  some integrable 1D  models, it is possible to determine the
parameters of the effective Hamiltonian nonperturbatively by
means of Bethe ansatz
\cite{RWA,CheianovPustilnik,ImambekovGlazman_PRL}.

The Hamiltonian of the TL model may be recast into the
Hamiltonian of free fermionic quasiparticles
\cite{Rozhkov05,Rozhkov06,Rozhkov08} having a linear spectrum:
\begin{eqnarray}
\tilde H_{1} = {\rm i}  v  \int dx \left[
        \colon
        \tilde \Psi^\dagger_{{\rm L}}(x)
        \nabla \tilde \Psi^{\vphantom{\dagger}}_{{\rm L}}(x)
        \colon
        -
        \colon
        \tilde \Psi^\dagger_{{\rm R}}(x)
        \nabla \tilde \Psi^{\vphantom{\dagger}}_{{\rm R}}(x)
        \colon
\right]. \label{H1} \eea Here   $\tilde \Psi^\dagger_{{\rm R
(L)}}(x)$ and $\tilde \Psi^{\vphantom{\dagger}}_{{\rm R (L)}}(x)$
are creation and annihilation operators for quasiparticles on the
right (left) branch, satisfying usual fermionic commutation
relations. Colons indicate the normal ordering with respect to
filled Fermi seas: for right (left) branch all states with
negative (positive) momenta are occupied.  The density of
quasiparticles $ \tilde \rho^{\vphantom{\dagger}}_{\rm R(L)} (x)=
\colon \tilde \Psi^\dagger_{{\rm R (L)}}(x)\tilde
\Psi^{\vphantom{\dagger}}_{{\rm R (L)}}(x) \colon  $ is simply
related to the density of  fermions in the TL model $
\rho^{\vphantom{\dagger}}_{\rm R(L)} (x)=\colon
\Psi^\dagger_{{\rm R (L)}}(x) \Psi^{\vphantom{\dagger}}_{{\rm R
(L)}}(x) \colon .$ Because the canonical transformation that
diagonalizes the TL Hamiltonian is a Bogoliubov rotation in the
space of particle-hole excitations, such a relation is linear, $
\rho^{\vphantom{\dagger}}_{\rm R}(x) +
\rho^{\vphantom{\dagger}}_{\rm L}(x)=
  K (\tilde \rho^{\vphantom{\dagger}}_{\rm R}(x) + \tilde
\rho^{\vphantom{\dagger}}_{\rm L}(x)).$ Fermionic operators  are
related to fermionic quasiparticles  using "string" operators
$\tilde F^{\dagger}_{\rm R (L)}(x)$ as (e.g. for right-movers)
\bea \Psi^\dagger_{{\rm R }}(x)= \tilde F^{\dagger}_{\rm R}(x)
\tilde \Psi^\dagger_{{\rm R }}(x), \; \tilde F^{\dagger}_{\rm
R}(x)= \exp{\left[ i \int^{x}dy \left( \delta_{+} \tilde
\rho^{\vphantom{\dagger}}_{\rm R}(y) + \delta_{-} \tilde
\rho^{\vphantom{\dagger}}_{\rm L}(y) \right)\right]}.
\label{FviaPsitilde}\eea Here we have introduced  parameters \bea
\frac{\delta_{+}}{2\pi} =
1-\frac{1}{2\sqrt{K}}-\frac{\sqrt{K}}{2}<0, \;
\frac{\delta_{-}}{2\pi}=\frac{1}{2\sqrt{K}}-\frac{\sqrt{K}}{2}.
\label{deltas}
 \eea
Using Eqs.~\ref{FviaPsitilde} and \ref{deltas}  together with
Eq.~\ref{H1}, one can obtain the usual results for Green's
function of the TL model \cite{Rozhkov05}.

If one wants to consider effects of nonlinearity, one has to
include terms that are less relevant in the renormalization group
sense into quasiparticle Hamiltonian. One such term is the
nonlinearity of the spectrum of quasiparticles:
 \bea
 \tilde H_{2}= \frac1{2m_*} \int dx \left(
        \colon
        (\nabla \tilde \Psi^\dagger_{{\rm L}})
        (\nabla \tilde \Psi^{\vphantom{\dagger}}_{{\rm L}})
        \colon
        +
        \colon
        (\nabla \tilde \Psi^\dagger_{{\rm R}})
        (\nabla \tilde \Psi^{\vphantom{\dagger}}_{{\rm R}})
        \colon
\right). \label{H2} \eea Here $m_{*}$ is the effective mass, which
can be related~\cite{Pereira_short} to low-energy properties as $
1/m_*= v/K^{1/2}\partial{v}/\partial{h}+
 v^2/(2K^{3/2})\partial{K}/\partial{h},$
 where $h$ is the chemical potential.

In principle, there is another term that needs to be included
together with Eq.~\ref{H2}: It amounts to interaction between
quasiparticles created by operators $\tilde\Psi^\dagger_{\rm
L,R}$. It can be shown \cite{Rozhkov06}, however, that in the
limit of small $p$ interactions between quasiparticles  are weak
and can be treated perturbatively, along the lines
of~\cite{Pustilnik2006Fermions,Khodas2006Fermions}. Perturbation
theory is valid as long as the interaction between the original
fermions (created by $\Psi^\dagger$) is short-ranged.
Interactions between quasiparticles are  responsible for weak
singularities in $S(p,\omega)$ near $\omega= v p\pm p^2/(2m_*),$
large-$\omega$ tails of $S(p,\omega)$
\cite{Pustilnik2006Fermions}, and for possible finite $\propto
p^8$ smearing \cite{Khodas2006Fermions} of some of the
singularities of $A(p,\omega).$   All these effects vanish as long
as one is interested in the scaling limit $ p\rightarrow 0,
\varepsilon \rightarrow \; \mbox{const},$ see SOM \cite{SOM} for
more detailed discussion.
 For models with
interactions decaying  as $\propto 1/x^2$ or slower, non-analytic
dependence of interactions on momentum becomes possible, and one
can not neglect interactions between quasiparticles. This can be
already seen from perturbative calculations
~\cite{Khodas2006Fermions}.

 The spectral function $A(p,\omega)$ gets modified  by the
spectrum nonlinearity in a  profound way because the dynamics of
the string operators operators $\tilde F^{\dagger}_{\rm
R(L)}(x,t)$ in Eq.~\ref{FviaPsitilde} becomes nonlinear.
Effective mass $m_*$ defines the energy scale $\sim p^2/(2m_*)$
near $\omega=v p$ where modifications from the TL model take
place. Because parameters $\delta_{\pm}$ defining $\tilde
F^{\dagger}_{\rm R(L)}(x,t)$ are universally related to $K,$ full
form of the crossover written in terms of a variable
$\varepsilon$ is a universal function of $K.$ Investigation of
the properties of crossover function $A(\varepsilon)$ is the main
subject of the present article.

Before proceeding to discuss the form of the universal crossover,
let us consider the main new features of $A(p,\omega)$ that arise
because of nonlinear spectrum. We find that in the vicinity of
each low-energy region $k\approx (2n+1) k_f$ spectral function
$A(p,\omega)$ has a power-law behavior near frequencies $\pm
\left[v p\pm p^2/(2m_*)\right],$ which is related to orthogonality
catastrophe phenomenon
\cite{Pustilnik2006Fermions,Khodas2006Fermions}: \bea A(p,\omega)
\propto {\rm const}+ \left|\frac{1}{\omega \pm
 \left(v p\pm \frac{p^2}{2m_*}\right)}\right|^{\mu}, \label{mudef}
\eea  and notations for $\mu$ are shown in Fig. \ref{Fig1}. Such
power-law behavior results from multiple low-energy particle-hole
excitations near left and right Fermi points, which are created
when  "high energy" fermion tunnels into the system.

 To be specific, let us focus on
the vicinity of $+k_f$ for $p>0$ and $ \omega>0.$  Because the
fermion that tunnels into the system has a momentum near $+k_f$
and energy of the system increases for $\omega>0,$ we need to
consider only the correlator $\bigl \langle
\Psi^{\vphantom{\dagger}}_{{\rm R}}(x,t)\Psi^{\dagger}_{{\rm
R}}(0,0) \bigr \rangle.$

Let us first discuss the exponent $\overline{\mu_{0,+}}$ at the
edge $\left| \omega - \left( v p + \frac{p^2}{2m_*}\right) \right|
\ll \frac{p^2}{2m_*}.$ To understand its origin, one has  to
understand the states that can be created
 by $\Psi_{\rm R}^\dagger,$ when the energy of the tunneling fermion is in the vicinity of the
edge. From energy and momentum conservation, such state is given
by a single fermionic quasiparticle with "large" momentum $\approx
p$ and multiple low-energy particle-hole excitations with momenta
much smaller then $p,$ as indicated in Fig.~\ref{Fig2}. Then  one
can neglect all other
states~\cite{Pustilnik2006Fermions,Khodas2006Fermions} and
project quasiparticle operators $\tilde
\Psi^{\vphantom{\dagger}}_{{\rm R}}(x)$ and $\tilde
\Psi^{\vphantom{\dagger}}_{{\rm L}}(x)$ onto narrow (of the width
much smaller than $p$) subbands $r, d,$ and $l$ as $  \tilde
\Psi^{\vphantom{\dagger}}_{{\rm R}}(x) \approx \tilde
\psi^{\vphantom{\dagger}}_{r}(x) + e^{i p x}\tilde d(x), \; \tilde
\Psi^{\vphantom{\dagger}}_{{\rm L }}(x) \approx \tilde
\psi^{\vphantom{\dagger}}_{l}(x).$

The effective Hamiltonian determining the evolution of these
states  is obtained by projecting $\tilde H_1 + \tilde H_2$ onto
subbands $r,l,$ and $d$ and linearizing the corresponding spectra:
 \bea
\tilde H_{r,l} ={\rm i} v  \int dx \left(
        \colon
        \tilde \psi^\dagger_{ l }(x)
        \nabla \tilde \psi^{\vphantom{\dagger}}_{{ l}}(x)
        \colon
        -
        \colon
        \tilde \psi^\dagger_{ r}(x)
        \nabla \tilde \psi^{\vphantom{\dagger}}_{ r}(x)
        \colon
\right), \label{Hrl}\\ \tilde H_{d}= \int dx \tilde d^{\dagger}
(x) \left[ v p+\frac{p^2}{2m_*}-i \left( v +\frac{p}{m_*}\right)
\nabla\right]\tilde d(x).\label{Hd} \eea The Green's function
factorizes as
 $  \propto e^{i p x} \langle\tilde d(x,t)\tilde
d^{\dagger}(0,0) \rangle_{\tilde H_d} \langle{\tilde
F}^{\vphantom{\dagger}}_{r} (x,t){\tilde F}^{\dagger}_{r}
(0,0)\rangle_{\tilde H_{r,l}}.$ To obtain string operators
${\tilde F}^{\vphantom{\dagger}}_{r},{\tilde F}^{\dagger}_{r}$
from Eq.~\ref{FviaPsitilde}, one should keep only $r$ and $l$
components of the density there.  The free-particle correlator
$\langle\tilde d(x,t)\tilde d^{\dagger}(0,0) \rangle_{\tilde H_d}$
 equals $  \propto e^{-i \left(v
p+\frac{p^2}{2m_*}\right) t} \delta \left[x- \right(v+
\frac{p}{m_*}\left) t\right],$  and string correlator can be
bosonized and evaluated~\cite{Giamarchibook} in a usual way
 as
$ \langle{\tilde F}^{\vphantom{\dagger}}_{r} (x,t){\tilde
F}^{\dagger}_{r} (0,0)\rangle_{\tilde H_{r,l}}|_{x=(v+
\frac{p}{m_*}) t}\propto
t^{-\left(\delta_-/(2\pi)\right)^2-\left(\delta_+/(2\pi)\right)^2}.
$ Taking Fourier transform of
$\langle\Psi^{\vphantom{\dagger}}_{\rm R}(x,t) \Psi^\dagger_{\rm
R}(0,0)\rangle,$ we obtain the universal exponent \bea
\overline{\mu_{0,+}}=
1-\left(\frac{\delta_-}{2\pi}\right)^2-\left(\frac{\delta_+}{2\pi}\right)^2.
\label{overlinemu+} \eea

Analogously, exponent $\underline{\mu_{0,+}}$ for $ \omega -
\left( v p - \frac{p^2}{2m_*}\right)\ll \frac{p^2}{2m_*}$ is
determined by configurations with one quasihole with the momentum
$\approx -p,$ two quasiparticles near right Fermi point, and
low-energy particle-hole excitations. One can again reduce the
problem to three-subband model and bosonize states near right and
left Fermi points. This way, one obtains the exponent
 \bea \underline{\mu_{0,+}}=
1-\left(\frac{\delta_-}{2\pi}\right)^2-\left(2-\frac{\delta_+}{2\pi}\right)^2<-3.
\label{underlinemu+} \eea

 New exponents given by Eqs.~\ref{overlinemu+} and \ref{underlinemu+} are
clearly different from the result for the TL model in Eq.~S4,
which corresponds to the exponent
$1-\left[\delta_-/(2\pi)\right]^2.$

Configurations responsible for the remaining exponents
$\overline{\mu_{0,-}}, \underline{\mu_{0,-}}$  consist of "high
energy" particle-hole excitation on the left branch, particle at
the right  Fermi point, and low-energy excitations on left and
right branches. Singularities near $k\approx (2n+1) k_f$ also
include $n$ low-energy particle-hole pairs with momentum $\approx
2 n k_f.$ All exponents can be obtained by using projections onto
three-subband models, and the results are summarized in Table
\ref{Table1}.

We now discuss the results for the universal crossover function
$A(\varepsilon)$  in the vicinity of $+k_f$ for $p, \omega>0$
[details of the derivations are available in SOM \cite{SOM}]. The
answer is defined by a universal function $D(y),$ determined only
by $\delta_+$ and normalized as $\int_{-1}^{1} D\left(y \right)
dy =1.$ By using $ D\left(y \right),$  spectral function can be
written as a convolution of contributions from the left and right
branches. Universal function $A(\varepsilon)$ in Eq.~\ref{edef}
is related to $D(y)$ as \bea A(\varepsilon)= \int_{-1}^{1} dy
D\left(y \right) \theta(\varepsilon -y )(\varepsilon
-y)^{(\frac{\delta_-}{2\pi})^2-1}. \label{conveq} \eea One can
analytically obtain limiting behavior of $D(y)$ for $y\rightarrow
\pm 1$ from Eqs.~\ref{overlinemu+} to \ref{conveq} as $ D(y)
\propto (1\mp y)^{d_{\pm}}  \; \mbox{for} \; y\rightarrow \pm1, $
where
 \bea d_+=\left(\frac{\delta_+}{2\pi}\right)^2-1,\; d_-= \left(2-\frac{\delta_+}{2\pi}\right)^2-1>3. \label{Dys}\eea

At moderate interaction strength, $\overline{\mu_{0,+}}>0$,
function $A(\varepsilon)$ diverges at $\varepsilon=1$. Then the
ratio of the prefactors above and below the singular line is
universal, \bea \lim_{|\delta \varepsilon|\rightarrow
0}\frac{A(1+|\delta \varepsilon|)}{A(1-|\delta
\varepsilon|)}=\frac{\Gamma\left((\frac{\delta_+}{2\pi})^2
\right)}{\Gamma\left((\frac{\delta_-}{2\pi})^2 \right)}
\frac{\Gamma\left(1-(\frac{\delta_+}{2\pi})^2
\right)}{\Gamma\left(1-(\frac{\delta_-}{2\pi})^2 \right)}.
\label{preexpratio} \eea

To evaluate $D(y)$ away from the edges, one should be able to
calculate the dynamics of chiral vertex operators \cite{SOM}. For
a nonlinear spectrum, this is a very nontrivial problem, the
analytic solution of which is not known. Similar  correlators have
attracted attention recently
~\cite{AbanovWiegmannPRL2,AbanovWiegmannrecent}, and their
connection to the nonlinear quantum shock wave dynamics and
nonlinear differential equations has been discussed. Although it
might be possible to proceed similarly for the evaluation of
$D(y),$ it is not clear whether nonlinear differential equations
obtained this way will have an analytic solution. We use an
alternative approach of~\cite{AL1, AL2}, which allows us to
develop a representation of $D(y)$ in terms of certain
determinants built of single-particle (rather than many-body)
states. These determinants can be evaluated numerically, which
practically solves the problem of finding $D(y).$ Representative
results for $D(y)$ and $A(\varepsilon)$ for $K=4.54$ are shown in
Fig. \ref{Fig3}.

The universal Hamiltonian given by Eqs.~\ref{H1} and \ref{H2} can
be also used to describe gapless bosonic and spin$-\frac{1}2$
systems away from particle-hole symmetric ground states. We
present main results on singularities of their dynamic response
functions in SOM \cite{SOM}.

 We  have constructed universal low-energy theory of a wide class of  interacting 1D
quantum liquids without  resorting to the simplifications of the
Tomonaga-Luttinger model accepted in the phenomenological
Luttinger liquid description. Unlike the latter, we keep the
nonlinear  dispersion relation of the fermions intact. The
replacement of the dispersion relation by a linear one,
$\omega=vp$, results in an  artificial introduction of Lorentz
invariance into the system. Although not affecting the low-energy
behavior of local properties (such as the local tunneling density
of states), the  introduced symmetry alters qualitatively the
predictions for the momentum-resolved quantities, such as the
spectral function. Keeping the nonlinearity allows us to find the
generic low-energy behavior of the dynamic response functions of a
system of interacting fermions, bosons, and spins. Possible
extensions of our theory should be able to describe the effects
of finite temperature, spin systems at particle-hole symmetric
points, systems with long-range interactions, and fermions with
spin.


\begin{scilastnote}
\item We thank  A. Kamenev and D. Abanin for  useful discussions. This work was
supported by US Department of Energy grant no. DE-FG02-08ER46482.
\end{scilastnote}

{\bf Supporting Online Material}\\
www.sciencemag.org \\
Materials and Methods\\
Fig. S1\\
Table S1\\
References

\begin{table}
\begin{tabular}{|c|c|}
\hline
 $ \overline{\mu_{n,+}}$ &
$1-\frac12\left(2n-(2n+1)\frac{\delta_++\delta_-}{2\pi}\right)^2-\frac12\left(\frac{\delta_+-\delta_-}{2\pi}\right)^2$
\\
$ \underline{\mu_{n,+}}$ &
$1-\frac12\left(2n+2-(2n+1)\frac{\delta_++\delta_-}{2\pi}\right)^2-\frac12\left(2-\frac{\delta_+-\delta_-}{2\pi}\right)^2$
 \\
  $ \overline{\mu_{n,-}}$ &
$1-\frac12\left(2n+2-(2n+1)\frac{\delta_++\delta_-}{2\pi}\right)^2-\frac12\left(\frac{\delta_+-\delta_-}{2\pi}\right)^2$
 \\
  $ \underline{\mu_{n,-}}$ & $1-\frac12\left(2n-(2n+1)\frac{\delta_++\delta_-}{2\pi}\right)^2-\frac12\left(2-\frac{\delta_+-\delta_-}{2\pi}\right)^2$ \\ \hline
\end{tabular}
\caption{ Universal exponents for spectral function. Notations
are indicated in Fig. \ref{Fig1}, and parameters $\delta_{\pm}$
defined by Eq. \ref{deltas} are functions of $K$ only. Note that $
\mu_{n,+}= \mu_{-n-1,-}, $ which follows from the $k\rightarrow
-k$ symmetry. \label{Table1}}
\end{table}

\begin{figure}
\includegraphics[width=14cm]{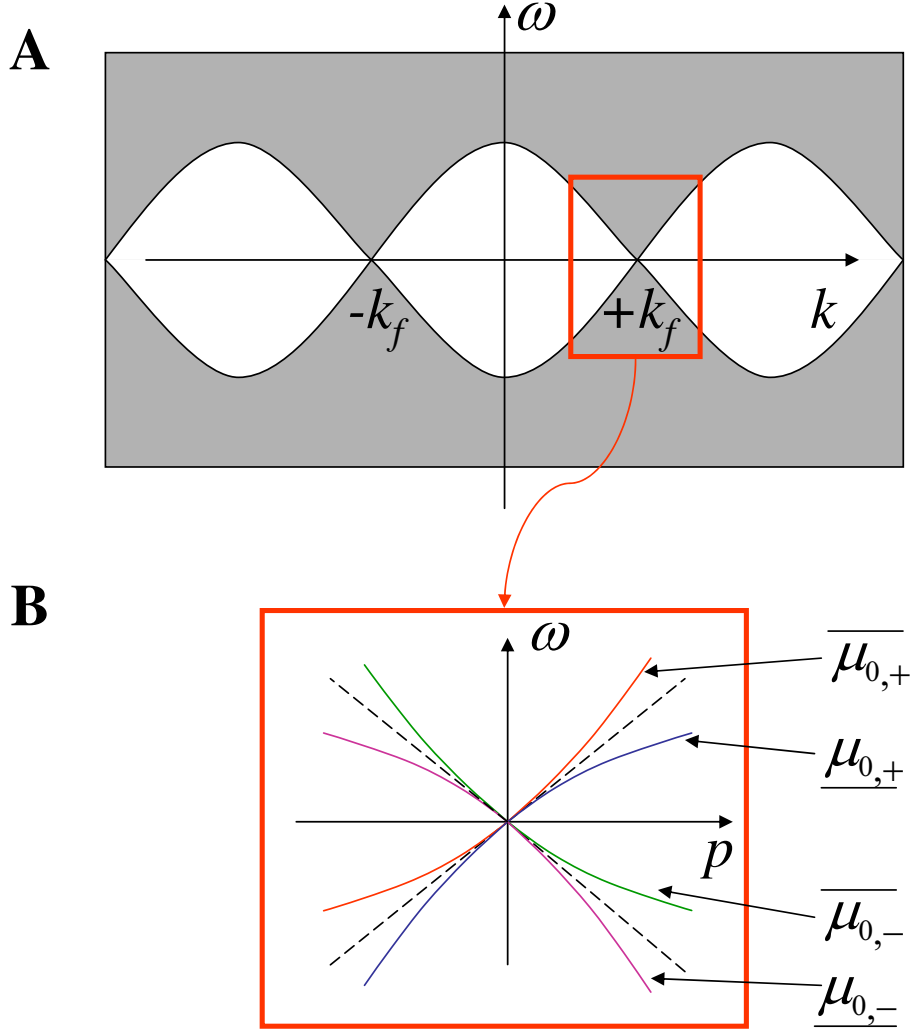}
\caption{\label{Fig1} Spectral function. {\bf (A)} Spectral
function $A(k, \omega)$ in momentum-energy plane. Shaded areas
indicate the regions where $A(k, \omega)\neq 0.$ The region with
$\omega>0$ corresponds to the particle  part of the spectrum, and
the region with $\omega<0$ corresponds to the hole part of the
spectrum. {\bf (B)} Close-up view of the vicinity of $k\approx
+k_f,$ where $p=k-k_f.$ Notations of $\mu$ indicate which
exponents presented in Table \ref{Table1} should be used in Eq.
\ref{mudef}. Notations for exponents near $k\approx(2n+1)k_f$ are
obtained by substituting corresponding $n$ instead of $n=0.$}
\end{figure}

\begin{figure}
\includegraphics[width=14cm]{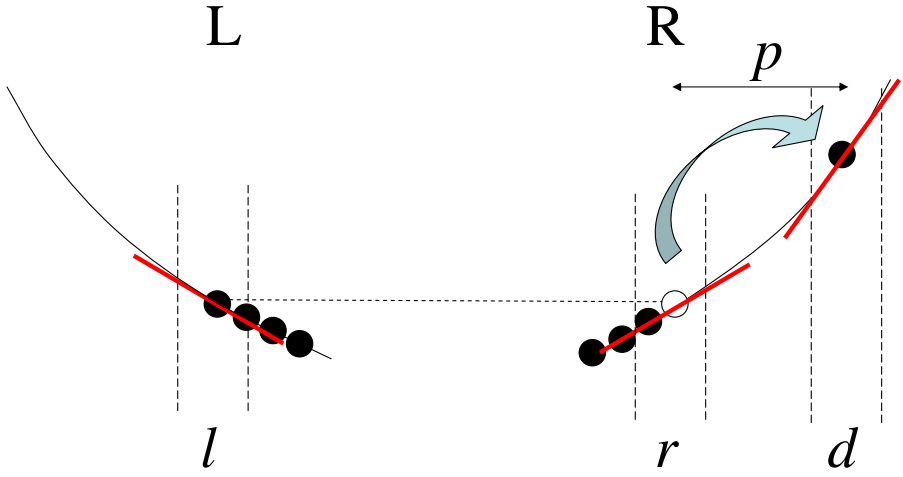}
\caption{\label{Fig2} Reduction to the effective Hamiltonian. We
show excitations contributing to the singularity at $\left|\omega
- \left( v p + \frac{p^2}{2m_*}\right)\right| \ll
\frac{p^2}{2m_*}.$ The Hamiltonian given by Eqs. \ref{H1} and
\ref{H2} is reduced to the three-subband model in Eqs.~\ref{Hrl}
and \ref{Hd}.}
\end{figure}

\begin{figure}
\includegraphics[width=14cm]{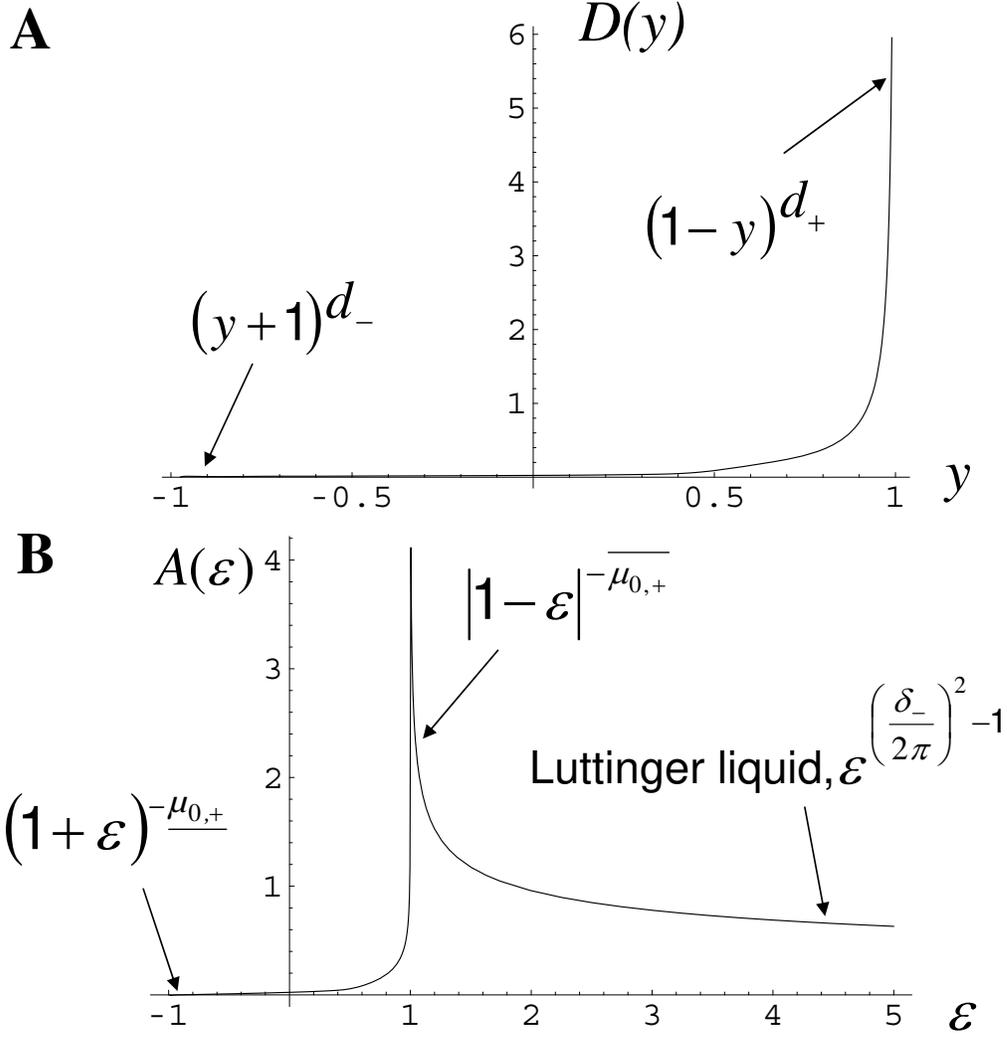}
\caption{\label{Fig3} Universal crossover. {\bf (A)} Universal
crossover function $D(y)$ for $K=4.54$ and the corresponding
values $\delta_+/(2\pi)=-0.3$ and $\delta_-/(2\pi)=-0.83;$ see Eq.
\ref{deltas}. Exponents $d_{\pm}$ defining the asymptotic
behavior at $y\rightarrow \pm 1$ are given by Eq.~\ref{Dys}. {\bf
(B)} Universal function $A(\varepsilon)$ for $K=4.54.$ Exponents
$\overline{\mu_{0,+}}$ and $ \underline{\mu_{0,+}}$ defining the
asymptotic behavior at $\varepsilon \rightarrow \pm 1$ are given
by Eqs.~\ref{overlinemu+} and \ref{underlinemu+}. The ratio of
prefactors determining the asymmetry of the singularity at
$\varepsilon=1$, see Eq. \ref{preexpratio}, equals $2.96$ for
$K=4.54.$
 }
\end{figure}

\clearpage \setcounter{equation}{0}%
\setcounter{figure}{0} \setcounter{table}{0}
\renewcommand{\theequation}{S\arabic{equation}}
\renewcommand{\thefigure}{S\arabic{figure}}
\renewcommand{\thetable}{S\arabic{table}}

\section*{Supporting Online Material}

\subsection*{Materials and Methods 1: Definitions of the dynamic correlation functions}

We are interested mostly in the zero-temperature spectral function
\bea A(k,\omega)= -\frac1{\pi}{\rm Im }G^{\rm ret}(k,
\omega),\label{Akw} \eea
 where retarded Green's function $G^{\rm
ret}(k,\omega)$ is defined by 
({\it S1}) \bea G^{\rm ret}(k, \omega)= -i \int\int dx dt e^{i
(\omega t -kx )} \times \bigl \langle
\Psi(x,t)\Psi^{\dagger}(0,0)+\Psi^{\dagger}(0,0)\Psi(x,t) \bigr
\rangle \theta(t), \label{SFdef} \eea and dynamic structure
factor (DSF) \bea S (p,\omega) = \int\!dx\,dt\,e^{i (\omega
t-px)}\, \bigl \langle \rho (x, t) \rho (0, 0) \bigr \rangle.
\label{DSFdef} \eea  Here $\Psi^{\dagger}(x,t) , \Psi(x,t)$ and
$\rho(x,t)$ are fermionic or bosonic creation, annihilation and
density operators, respectively. Energy $\omega$ is measured
respective to the chemical potential, so $A(k,\omega)$ for
$\omega>0 \;(\omega<0)$ describes the response of the system to
an addition of an extra particle (hole).

\subsection*{Materials and Methods 2: Universal crossover}
In this section we present the details of the derivations the
universal crossover function $A(\varepsilon)$  in the vicinity of
$+k_f$ for $p, \omega>0.$ Before proceeding to the case of the
nonlinear spectrum, let us present the derivation of a
conventional result for the Tomonaga-Luttinger model, \be
A(p,\omega)\propto (\omega -
vp)^{\frac{1}{4}\left(K+\frac{1}{K}-2 \right)-1} \theta (\omega -
vp) \label{Ape}, \ee  which allows for an easy generalization to
the nonlinear case.

Retarded Green's function for fermions near $+k_f$ can be written
({\it S2}) as a product of two terms,  determined by left and
right Fermi points. Due to linear spectrum, they depend on
combinations $ v t+x$ and $v t-x,$ respectively:
 \bea G^{\rm ret}_{\rm R}(x,t) \propto -i \frac{\theta(v t+x)}{(i(v t+x)+0)^{(\frac{ \delta_-}{2\pi})^2} }
 \frac{\theta( v t-x)}{(i(v t-x)+0)^{(\frac{ \delta_-}{2\pi})^2+1}}. \eea
 If one defines $L(x,t)$ and $R(x,t)$  as
\bea  L(x,t)\propto\frac{1}{(i(v t+x)+0)^{(\frac{
\delta_-}{2\pi})^2} }, \;R(x,t)\propto\frac{1}{(i(v
t-x)+0)^{(\frac{ \delta_-}{2\pi})^2+1}}, \eea then imaginary part
of Fourier transform of $G^{\rm ret}_{\rm R}(x,t)$ can be
represented as a convolution of two Fourier transforms of $L(x,t)$
and $R(x,t):$ \bea {\rm Im } G^{\rm ret}_{\rm R}(p,\omega)= -i
\int\frac{d\tilde p}{2\pi}\frac{d\tilde \omega}{2\pi} L(p-\tilde
p,\omega-\tilde \omega) R(\tilde p, \tilde \omega),
\label{convolution}\eea where real positive functions
$R(p,\omega)$ and $L(p,\omega)$ are equal to (up to a positive
cut-off dependent prefactor) \bea L(p,\omega) \propto
\delta(\omega+ v p)\theta(\omega- v p)
(\omega -  v p )^{(\frac{ \delta_-}{2\pi})^2-1}, \label{Lkw}\\
R(p,\omega)\propto \delta(\omega- v p) \theta(\omega+ v p)(\omega
+  v p)^{(\frac{ \delta_-}{2\pi})^2} . \label{Rkw}\eea Using
Eq.~\ref{Akw} spectral function $A(p,\omega)$ can be written as
\bea A(p,\omega)=\frac{1}{\pi} \int\frac{d\tilde p
}{2\pi}\frac{d\tilde \omega}{2\pi} L(p-\tilde p,\omega-\tilde
\omega)R(\tilde p, \tilde \omega). \label{Aconv} \eea Physically,
$A(p,\omega)$ describes the  probability of tunneling of a
fermion with total energy $\omega$ and momentum $p.$
Eqs.~\ref{Lkw}-\ref{Rkw} then mean that excitations which are
created on the right (left) branch should lie on the respective
mass shell and have positive (negative) momenta. From energy and
momentum conservation laws, nonzero contribution to $A(p,\omega)$
for $\omega>v p$ comes only from a single point  in the integral
in Eq.~\ref{Aconv}, which correspond to the following arguments
of functions $R(p_{\rm R},\omega_{\rm R})$ and $L(p_{\rm
L},\omega_{\rm L})$ in the integrand: \bea
 \omega_{\rm R}=v p_{\rm R}=\frac{(\omega + v p)}2, \label{rb}\\
\omega_{\rm L}=- v p_{\rm L}=\frac{(\omega - v p)}2. \label{lb}
 \eea
  Since only $L( p_{\rm L}, \omega_{\rm L})$ is singular for
$\omega \rightarrow v p,$ for the Tomonaga-Luttinger model only
the contribution due to the shake-up of low-energy excitations
near the left Fermi point controls the exponents at $\omega
\approx v p.$ For nonlinear spectrum, shake-up contributions from
both left and right Fermi points determine the exponents, see e.g.
Eq.~9.

 For nonlinear spectrum, $ G^{\rm ret}_{\rm R}(x,t)$ can still be
represented as a product of two terms determined by left and right
Fermi points. The primary modification which takes place is that
delta-functions in Eqs. \ref{Lkw}-\ref{Rkw} get broadened.
 Indeed, for nonlinear spectrum the total  momentum of several left (right)-moving quasiparticles doesn't completely
define their total energy, and the latter is allowed to vary up to
$ \pm p_{\rm L(R)}^2/2m_{*}.$ However, if one is interested in the
scaling behavior of $A(p,\omega)$ for  \bea \omega- v p =
\varepsilon \frac{p^2}{2m_{*}}, \label{regiondef}\eea one can
neglect the finite width of function $L,$ since according to
Eqs.~\ref{lb} and \ref{regiondef} it is of the order $p_{\rm L}^2
\propto p^4,$ which vanishes in the scaling limit $ p\rightarrow
0, \varepsilon \rightarrow \mbox{const}.$ Broadening of function
$R,$ on the other hand, is important. As a result of it the
momentum and energy on each branch can vary on the order $\sim
p^2/(2m_* v)$ and  $\sim p^2/(2m_*),$ respectively, around the
values of Eqs.~\ref{rb},\ref{lb}.

To characterize the broadening of  delta-function in
Eq.~\ref{Rkw},  let us introduce a dimensionless positive
function $D(y)$  defined by \bea R(p, vp +
 y \frac{p^2}{2m_{*}})\propto D(y).\eea
Since for total momentum $p$ on the right branch $ v p\pm
p^2/(2m_*)$ is the highest (lowest) possible energy of a set of
quasiparticles, $D(y)\neq 0$ only if $y\in (-1,1).$ Universal
function $D(y)$ is determined only by $\delta_+,$ and  we choose
it to be normalized as \bea \int_{-1}^{1} D\left(y \right) dy
=1.\eea

We now discuss how to reduce the evaluation of $D(y)$ to a
single-particle problem and solve it numerically. We use periodic
boundary conditions on  a circle of length $L.$ Since from now on
we will be dealing only with fermions at the right branch, we
drop index $\rm R$ for clarity of notations, and set $m_*=1/2.$

The chiral vertex correlation function which determines $D(y)$ can
be written as \bea R'(x',t)= \langle  e^{i \tilde H_2 t} \tilde
\Psi(x') \exp{\left[ -i \int^{x'} dy \delta_{+} \tilde
\rho(y)\right]}e^{-i \tilde H_2 t} \exp{\left[
 i \int^{0}dy \delta_{+}\tilde
\rho(y) \right]} \tilde \Psi^{\dagger}(0)\rangle,\label{Rdef}\eea
where one has to average over filled Fermi sea on the right
branch. In Eq.~\ref{Rdef} we took into account the effect of the
linear-spectrum Hamiltonian $\tilde H_1,$  Eq.~2, by shifting
$x'=x-vt.$ Universal function $D(y)$ is determined by the Fourier
transform of $R'(x',t)$ as \bea R'(p,t)=\int  dx' e^{- i p x'}
R'(x',t) \propto \int_{-1}^{1} e^{i p^2 y t} D(y) dy.
\label{RvsD} \eea In momentum space, $R'(x',t)$ can be written as
 \bea R'(x',t) =
\prod_{k<0}  e^{ i k^2 t } \sum_{p, p'} e^{ i p' x'} \langle
\tilde \Psi^{\vphantom{\dagger}}_{p'}  e^{\cal B } e^{-i {\cal H}
t} e^{\cal A}\tilde \Psi^{\dagger}_p \rangle, \label{Rxtmom} \eea
where operators $\cal A,\cal B$ and ${\cal H}$ act in a many-body
Hilbert space as
 \bea {\cal A}= -\frac{
\delta_+}{2\pi} \sum_{p\neq p'} \frac{2\pi}{L (p-p')}\tilde
\Psi^{\dagger}_p\tilde
\Psi^{\vphantom{\dagger}}_{p'}=\sum_{p,p'}\hat a_{p,p'} \tilde
\Psi^{\dagger}_p \tilde \Psi^{\vphantom{\dagger}}_{p'} ,\\ {\cal
B}= \frac{\delta_+}{2\pi}\sum_{p\neq p'} \frac{2\pi e^{- i (p-p')
x}}{L (p-p')} \tilde \Psi^{\dagger}_p \tilde
\Psi^{\vphantom{\dagger}}_{p'}=\sum_{p,p'}\hat b_{p,p'} \tilde
\Psi^{\dagger}_p \tilde \Psi^{\vphantom{\dagger}}_{p'}, \\ {\cal
H}= \sum_{p} p^2 \tilde \Psi^{\dagger}_p\tilde
\Psi^{\vphantom{\dagger}}_{p}=\sum_{p,p'}\hat h_{p,p'} \tilde
\Psi^{\dagger}_p \tilde \Psi^{\vphantom{\dagger}}_{p'},
 \eea while
$\hat a, \hat b $ and $\hat h$ are  matrices acting in a
single-particle Hilbert space. We introduce the density matrix
\bea \hat \rho= \frac1{Z}e^{- \sum_p \lambda_p \tilde
\Psi^{\dagger}_p \tilde \Psi^{\vphantom{\dagger}}_p},
e^{-\lambda_p}=\frac{n_p}{1-n_p}, \eea where $n_p$ is the
occupation number of mode $p,$ which we will be set to
$\theta(-p)$ at the end of the calculation. Then Eq. \ref{Rxtmom}
can be written as a trace over full many-body Hilbert space as
\bea R'(x',t) = \prod_{k<0} e^{ i k^2 t } \sum_{p, p'} e^{ i p'
x'} \mbox{Tr}\left(\tilde \Psi^{\vphantom{\dagger}}_{p'} e^{\cal
B} e^{- i {\cal H} t} e^{\cal A}\tilde \Psi^{\dagger}_p \hat \rho
\right). \nonumber \eea We use the relation \bea \tilde
\Psi^{\dagger}_p \hat \rho =\hat  \rho e^{\lambda_p} \tilde
\Psi^{\dagger}_p, \eea which effectively restricts summation to
$p>0,$ and the cyclic property of a trace to get
 \bea  R'(x',t) =
\prod_{k<0} e^{ i k^2 t } \sum_{p>0, p'} e^{ i p'
x'}\mbox{Tr}\left( e^{\cal B} e^{-i {\cal H} t} e^{\cal A} \hat
\rho \tilde \Psi^{\dagger}_p \tilde \Psi^{\vphantom{\dagger}}_{p'}
\right). \nonumber \eea This trace over many-body  Hilbert space
can be written via determinants of matrices acting in a
single-particle Hilbert space as~({\it S3-S7})
\bea R'(x',t) = \prod_{k<0} e^{ i k^2 t }
\sum_{p, p'} e^{ i p' x'} (1-n_p) \times \nonumber \\
\mbox{Det}(\hat I-\hat n+e^{\hat b} e^{-i \hat h t}e^{\hat a}
\hat n) \left( \hat n + e^{-\hat a} e^{i \hat h t}e^{- \hat b}
(1-\hat n)\right)^{-1}_{p',p}, \label{Rxtfinal}\eea where $\hat
n$ is a diagonal matrix with $n_p$ on the diagonal.

 To extract $D(y),$ one needs to evaluate $R'(p,t)$
defined by Eq.~\ref{RvsD} at times \bea t_\alpha= \frac{\pi
\alpha}{p^2}\eea for integer $\alpha.$ According to
Eq.~\ref{RvsD}, it corresponds to Fourier series coefficient of
$D(y):$ \bea R'(p,t_\alpha) \propto D_\alpha =\int^1_{-1} e^{i \pi
\alpha y} D(y) dy. \eea Function $D(y)$ can be written in terms of
$D_{\alpha}$ as \bea D(y) =\frac{D_0}{2}+\sum_{\alpha=1}^{\infty}
\mbox{Re}\left[D_\alpha e^{-i \pi \alpha y}\right]
\label{Dsum}\eea
 Since  for small enough $\delta_+$ function $D(y)$ has a singularity
given by Eq.~12 for $y\rightarrow 1,$  one expects \bea
D_{\alpha}\propto (-1)^{\alpha}
\alpha^{-(\frac{\delta_+}{2\pi})^2} , \; \mbox{for} \;\alpha
\rightarrow \infty. \label{Dasymp}\eea

 We
evaluate $R'(x',t_{\alpha})$ for various $x'$  using
finite-dimensional Hilbert space of the size up to $\sim 300,$ and
obtain its Fourier transform $R'(p,t_{\alpha}).$ Due to periodic
boundary conditions and finite size effects, asymptote given by
Eq.~\ref{Dasymp} doesn't hold for  largest $\alpha$ obtained
numerically. However, we find an excellent fit
 for sufficiently large $\alpha$ as \bea  D_{\alpha} \propto
e^{i c \alpha}(-1)^{\alpha}\left(\sin{\frac{\alpha}{\tilde
\alpha}}\right)^{-\gamma}. \label{Dfit}\eea
 Exponent $\gamma$ obtained using such fitting
procedure equals $(\frac{\delta_+}{2\pi})^2$ with a very high
accuracy. In Eq.~\ref{Dfit}, $c\ll 1$ accounts for a possible
shift of the frequency, while  finite $\tilde \alpha$ accounts for
finite size effects due to finite $kL/(2\pi)$ considered. We
remove the latter effects by smoothly substituting $D_\alpha$ in
Eq.~\ref{Dsum} by \bea  D_{\alpha} \propto e^{i a
\alpha}(-1)^{\alpha}\left(\frac{\alpha}{\tilde
\alpha}\right)^{-\gamma}, \label{Dsubst}\eea for $\alpha$ larger
then some intermediate $\alpha_* \ll \tilde  \alpha,$ and keeping
numerical results for smaller $\alpha.$ The sum in Eq.~\ref{Dsum}
with $D_{\alpha}$ given by Eq.~\ref{Dsubst} can be written in
terms of polylogarithmic functions.  We sum the contributions to
Eq.~\ref{Dsum} coming from large $\alpha$ using polylogarthmic
functions, while for smaller $\alpha$ we use numerical results.

 The procedure to extract $D(y)$
described above is very robust, and is not sensitive to particular
choice of parameters at the accuracy of about $\sim2\%$  or
$0.02,$ whichever is larger, for data presented in Fig. 3. As an
independent check, it reproduces the result $D(-1)\approx 0$ very
well. The correct value of the exponent $d_->3$ characterizing
the asymptote $D(y\to -1)$ is harder to reproduce.

\subsection*{Materials and Methods 3: Bosonic and spin systems}

The universal Hamiltonian given by Eqs.~2,5 can be also used to
describe gapless bosonic and spin$-\frac{1}2$ systems away from
particle-hole symmetric ground states. The only modification is
the existence of an additional Jordan-Wigner "string" operator in
the expression, in terms of fermions, for creation operator of
bosons $\Psi^{\dagger}_{\rm B}$ and for spin raising operator
$S^{+},$ respectively. Here we will only discuss the
singularities.

For bosons, existence of new singularities in response functions
of an integrable Lieb-Liniger~({\it S8})
model has been pointed out recently~({\it S9})
and in low-energy regime expressions for the exponents  in terms
of the Luttinger parameter $K$ have been obtained.  Same
exponents can be obtained using the methods of the current
article, which demonstrates their universality. Exponents in the
vicinity of the low energy region $k\approx 2\pi n k_f$ are
summarized in Table \ref{TableS1}, and notations of bosonic
exponents $\mu^{b}$ are indicated in Fig. \ref{FigS1}.

For spin$-\frac12$ systems, our results apply generally for the
following antiferromagnetic ($J>0$) Hamiltonian in a finite
magnetic field $h$: \bea H=J\sum_{i} S^{x}_{i}S^{x}_{i+1}+
S^{y}_{i}S^{y}_{i+1}+\sum_{i>j} V_{i-j} S^{z}_{i}S^{z}_{j} -
h\sum_{i} S^z_i.\nonumber\eea Here $S^{x,y,z}_i$ are
spin$-\frac12$ operators, and $V_i$ are assumed to decay faster
than $1/i^2$, and to be small enough so that the system is
gapless. We require finite magnetic field, since otherwise due to
particle-hole symmetry one quite generally has $m_*=\infty.$ In
this case the regime discussed in present article disappears, as
has been pointed out recently~({\it S10})
for an integrable XXZ model. While for fermionic systems without a
lattice one expects $m_*>0,$  it is not necessarily the case for
spins on a lattice. This can change the relative position of the
singularities compared to Fig.~1. For small enough interactions
one expects $m_*>0 (m_*<0)$ for negative (positive) magnetic
field $h,$ although for small enough magnetic fields interactions
can reverse the sign of $m_*,$ see e.g.~({\it S11}).

We will be interested in transverse dynamic spin structure factor,
defined by \bea S^{-+} (k,\omega) = \sum_{j}e^{-i k j}\int
\,dt\,e^{i \omega t}\, \bigl \langle S_j^{-} (t) S_{0}^{+} (0)
\bigr \rangle. \eea It is nonvanishing at low energies in the
vicinity of $k=\pi,$  {as long as the spin chain remains gapless}
~({\it S2,S12}).  Generalization of the approach described earlier
leads to
 \bea S^{-+} (k, \omega) \propto
 {\rm const}+ \left|\frac{1}{\omega -
 \left(v |k-\pi| \pm \frac{(k-\pi)^2}{2m_*}\right)}\right|^{\pm \frac{1}{\sqrt{K}}-\frac{1}{2K}},\nonumber\eea
  for $|k-\pi|\ll 1$. Here we have already expressed parameters $\delta_{\pm}$ as
 functions of $K$ using Eq.~4.

\subsection*{Materials and Methods 4: Limits of applicability}
In this section we discuss the limitations of  and leading
corrections to the universal results. One regime, when universal
results are not applicable has been already pointed out above,
and corresponds to $m_*=\infty.$ Such situation generically
arises for spin-$\frac12$ system at half-filling, when leading
correction to spectrum nonlinearity starts from terms $\propto
p^3.$ If leading $\propto p^2$ curvature of the spectrum is
non-vanishing, then our results quite generically apply for \bea
\frac{p}{k_f}\ll 1. \label{vallim} \eea
We show below that leading corrections to universal results are
suppressed in powers of this small parameter. To be specific, we
consider the modifications of singularities of fermionic
$A(p,\omega)$ for $p>0, \omega>0$ in the vicinity of $+k_f.$

 There are two types of terms which modify
the universal Hamiltonian. One type of terms corresponds to higher
order corrections to single-particle spectrum. Such terms merely
shift the positions of the singularities, but do not change the
exponents. Indeed, reduction to three-subband model only requires
velocity of $d-$particle to be different from $v.$ Since this
happens already for leading spectrum nonlinearity $\propto p^2,$
higher order curvature of the spectrum doesn't directly affect
$\overline{\mu_{0,+}}$ and $\underline{\mu_{0,+}}.$

 Second type of terms corresponds to irrelevant interactions
between fermionic quasiparticles. One such term, which has the
same scaling dimension as spectrum nonlinearity, is given by~({\it
S11,S13,S14})
\bea \tilde H_{\rm int}'= - i\tilde g' \int dx \left( \tilde
\rho_{\rm R} \left[\colon \tilde \Psi^\dagger_{\rm L} \nabla
\tilde \Psi^{\vphantom{\dagger}}_{\rm L} \colon
    - \colon \nabla \tilde \Psi^\dagger_{\rm L}\tilde \Psi^{\vphantom{\dagger}}_{\rm L}\colon
    \right]
-\tilde \rho_{\rm L} \left[\colon \tilde \Psi^\dagger_{\rm R}
\nabla \tilde \Psi^{\vphantom{\dagger}}_{\rm R}\colon
    - \colon \nabla \tilde \Psi^\dagger_{\rm R}\tilde
    \Psi^{\vphantom{\dagger}}_{\rm R}\colon\right]
    \right)
\label{Hint'}, \eea where $g'$ can be related~({\it S11,S14})
to low energy properties similar to $1/m^*,$ and generally these
quantities are of the same order of magnitude. Effect of such
interactions on e.g. $\overline{\mu_{0,+}}$ can be understood
using the methods of Refs.~({\it S15,S16}).
Indeed, after projection to three-subband model interactions lead
to modification of phase shift $\delta_-$ of the order \bea \Delta
\delta_- \sim \frac{g' p}{v_d -(-v)}\sim \frac{p}{k_f} \ll 1, \eea
and thus lead to small corrections to $\overline{\mu_{0,+}},
\underline{\mu_{0,+}}.$ Here $v_d$ is the velocity of particle
$d,$ which equals \bea v_d=v+ \frac{p}{m^*}. \eea Less relevant
interactions between left and right branches lead to even
stronger suppressed corrections to the exponents. One should
note, that presence of finite $g'$ also leads to $\propto p^8$
smearing of the singularity $\overline{\mu_{0,+}}$ for $\omega>0,$
while singularity $\underline{\mu_{0,+}},$ being a singularity at
a true kinematic border, remains intact~({\it S16}).

Another irrelevant interaction term which modifies the exponents
in linear order over $p/k_f$ arises due to momentum dependence of
interactions on the same branch, \bea \tilde H_{\rm int}''= \int
dp V(p) \left(\tilde \rho_{\rm L}(p) \tilde  \rho_{\rm L}(-p)+
\tilde \rho_{\rm R}(p) \tilde \rho_{\rm R}(-p)\right). \eea
Interaction $V(p)$ should vanish for $p \rightarrow 0,$ and have
the symmetry property $V(p)=V(-p).$ If one assumes that
 $V(p)$ is regular, then its expansion starts from the term
 $\propto p^2,$ and correction to the phase shift $\delta_+$ is of
 the order
 \bea
 \Delta \delta_+ \sim \frac{V(p)}{v_d -v}\sim \frac{p}{k_f} \ll 1.
 \eea
However, for interactions that decay as or slower than $\propto
1/x^2,$ momentum dependent part of $V(p)$ doesn't have to be
regular. Indeed, for models with interactions decaying as
$\propto 1/x^2,$ one has  $V(p)\sim |p|,$ which leads to finite
$\Delta\delta_+$ in the limit $p\to 0$, and a finite modification
of the universal exponents. Thus our universal results do not
apply to Haldane-Shastry~({\it S17,S18})
or Calogero-Sutherland~({\it S19})
models. In the case of the latter, this can be seen from explicit
calculations~({\it S16,S20}).

Finally, we note that predictions of universal Hamiltonian for
$S(p,\omega)$ for small $p$ can be checked using sum rules.
Universal Hamiltonian  given by Eqs.~2,5 predicts~({\it S21})
 that $S(p,\omega)$ at any interaction strength
approaches the form characteristic for free  fermions \bea
S(p,\omega)= \frac{m_*
K}{p}\theta(\frac{p^2}{2m_*}-|\omega-v|p||), \label{DSFuniv} \eea
once $p$ becomes small enough. One can check, that for
Galilean-invariant systems this result explicitly satisfies f-sum
rule~({\it S22}) 
\bea \int \omega S(p,\omega)\frac{d\omega}{2\pi}= \frac1{2\pi}vK
p^2=\frac{np^2}{2m}.\eea In addition, for all systems
compressibility sum rule (see e.g. Eq.~7.52 of Ref.~({\it S22})
) is also satisfied: \bea \lim_{p\rightarrow0} \int \frac1{\omega}
S(p,\omega)\frac{d\omega}{2\pi}=\frac{K}{2\pi
v}=\frac12\frac{\partial n}{\partial \mu}, \eea where in last
equation we have used relation of $K$ to compressibility,
see e.g. Eq.~2.59 of Ref.~({\it S2}). 

\begin{table}
\begin{tabular}{|c|c|}
\hline
 $ \overline{\mu^b_{n,+}}$ &
$1-\frac12\left(2n-1-2n\frac{\delta_++\delta_-}{2\pi}\right)^2-\frac12\left(\frac{\delta_+-\delta_-}{2\pi}\right)^2$
\\
$ \underline{\mu^b_{n,+}}$ &
$1-\frac12\left(2n+1-2n\frac{\delta_++\delta_-}{2\pi}\right)^2-\frac12\left(2-\frac{\delta_+-\delta_-}{2\pi}\right)^2$
 \\
  $ \overline{\mu^b_{n,-}}$ &
$1-\frac12\left(2n+1-2n\frac{\delta_++\delta_-}{2\pi}\right)^2-\frac12\left(\frac{\delta_+-\delta_-}{2\pi}\right)^2$
 \\
  $ \underline{\mu^b_{n,-}}$ & $1-\frac12\left(2n-1-2n\frac{\delta_++\delta_-}{2\pi}\right)^2-\frac12\left(2-\frac{\delta_+-\delta_-}{2\pi}\right)^2$ \\ \hline
\end{tabular}
\caption{Universal exponents for bosonic spectral function.
Notations are indicated in Fig. \ref{FigS1}, and parameters
$\delta_{\pm}$ defined by Eq.~4 are functions of $K$ only. Note
that $ \mu^b_{n,+}= \mu^b_{-n,-}, $ which follows from the
$k\rightarrow -k$ symmetry. \label{TableS1}}
\end{table}


\begin{figure}
\includegraphics[width=14cm]{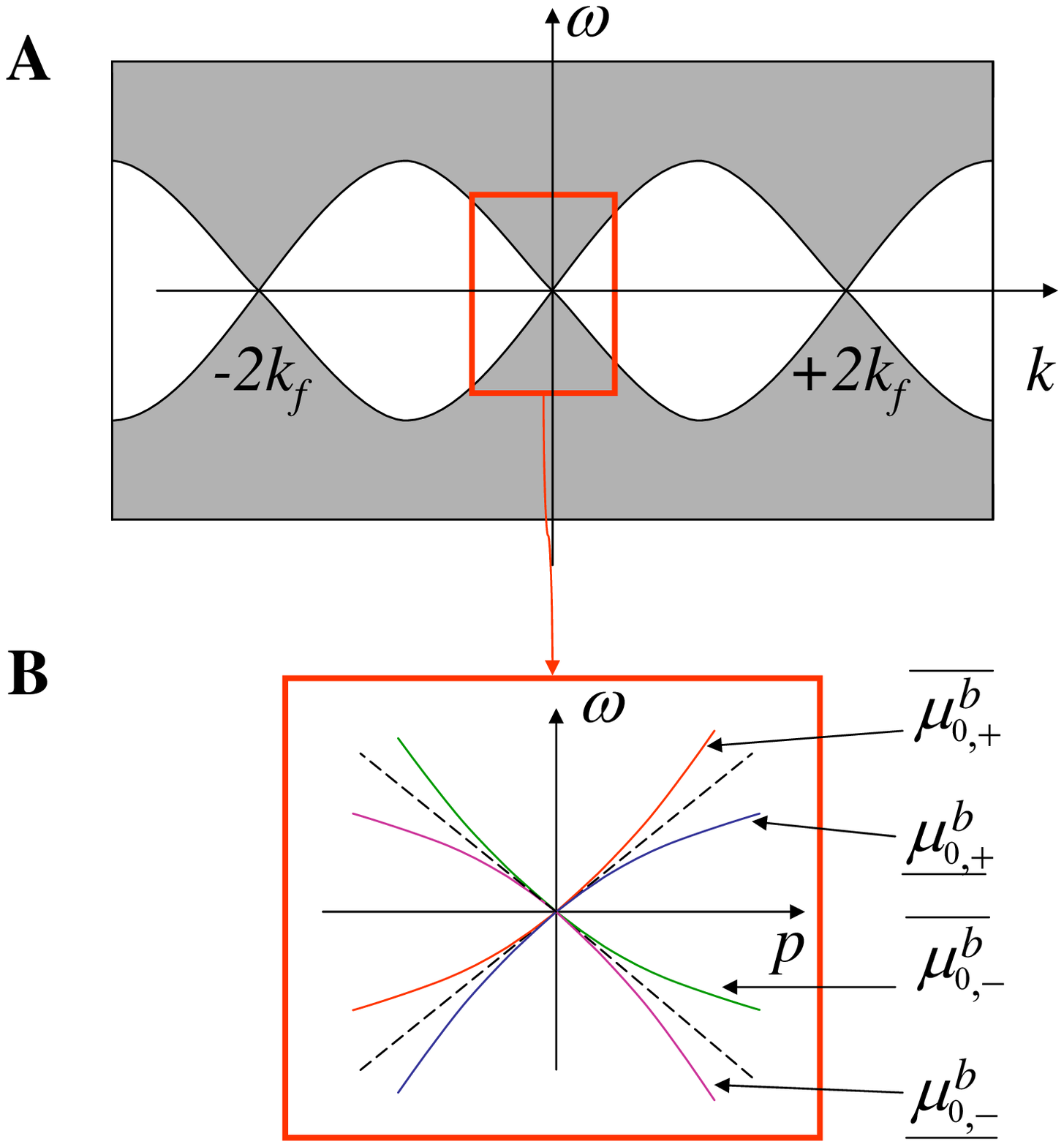}
\caption{\label{FigS1} Bosonic spectral function. {\bf (A)}
Spectral function $A(k, \omega)$ in momentum-energy plane. Shaded
areas indicate the regions where $A(k, \omega)\neq 0.$ The region
with $\omega>0\; (\omega<0)$ corresponds to the particle (hole)
part of the spectrum. {\bf (B)} Close-up view of the vicinity of
$k\approx 0,$ where $p=k.$ Notations of $\mu$ indicate which
exponents presented in Table \ref{TableS1} should be used in
Eq.~6. Notations for exponents near $k\approx 2n k_f$ are obtained
by substituting corresponding $n$ instead of $n=0.$}
\end{figure}

\section*{References}

\begin{itemize}

\item[S1.]A.A. Abrikosov, L.P. Gorkov, I.E. Dzyaloshinski,
 {\it Methods of Quantum Field Theory in Statistical Physics} (Dover, New York, 1963).
\item[S2.]T. Giamarchi, {\it Quantum Physics in One Dimension}
(Oxford Univ. Press, New York,  2004).
\item[S3.]R. Blankenbecler, D.J. Scalapino, R.L. Sugar, {\it Phys. Rev. D}
{\bf 24}, 2278 (1981).
\item[S4.]J.E. Hirsch, {\it Phys. Rev. B} {\bf 31}, 4403 (1985).
\item[S5.]I. Klich,  in {\it Quantum Noise in Mesoscopic Systems,}  Yu. V. Nazarov, Ed. (Kluwer, Dordrecht,
2003).
\item[S6.]D.A. Abanin, L.S. Levitov, {\it Phys. Rev. Lett.} {\bf 93}, 126802
(2004).
\item[S7.]D.A. Abanin, L.S. Levitov, {\it Phys. Rev. Lett.}  {\bf 94}, 186803
(2005).

\item[S8.]E.H. Lieb, W. Liniger, {\it Phys. Rev.} {\bf 130}, 1605
(1963).

\item[S9.]A. Imambekov,  L.I. Glazman,
{\it  Phys. Rev. Lett.}  {\bf 100}, 206805 (2008).

\item[S10.]V.V. Cheianov,   M. Pustilnik, {\it
Phys. Rev. Lett.}  {\bf 100}, 126403 (2008).

\item[S11.]R.G. Pereira {\it et al.},  {\it J. Stat. Mech.}  P08022 (2007).
\item[S12.]R. Chitra and T. Giamarchi, {\it Phys. Rev. B} {\bf 55}, 5816 (1997).

\item[S13.]A.V. Rozhkov, {\it Phys. Rev. B } {\bf 74}, 245123 (2006).

\item[S14.]R.G. Pereira {\it et al.},  {\it Phys. Rev. Lett.} {\bf 96}, 257202
(2006).
\item[S15.]M. Pustilnik, M. Khodas, A. Kamenev,  L.I. Glazman, {\it Phys. Rev. Lett.} {\bf 96}, 196405 (2006).
\item[S16.]M. Khodas, M. Pustilnik, A. Kamenev,  L.I. Glazman, {\it Phys. Rev. B}  {\bf 76}, 155402
(2007).

\item[S17.]F.D.M. Haldane, {\it Phys. Rev. Lett.} {\bf 60}, 635 (1988).
\item[S18.]B.S. Shastry, {\it Phys. Rev. Lett.} {\bf 60}, 639 (1988).
\item[S19.]B. Sutherland, {\it Beautiful Models} (World Scientific, Singapore, 2004).
\item[S20.]M. Pustilnik, {\it Phys. Rev. Lett.} {\bf 97}, 036404 (2006).
\item[S21.]A.V. Rozhkov, {\it Phys. Rev. B } {\bf 77}, 125109 (2008).
\item[S22.]L.P. Pitaevskii, S. Stringari, {\it Bose-Einstein Condensation} (Oxford Univ. Press, New York, 2003).

\end{itemize}

\end{document}